\documentstyle[12pt,epsf]{article}

\begin{document}

\baselineskip 6mm
\renewcommand{\thefootnote}{\fnsymbol{footnote}}


\newcommand{\nc}{\newcommand}
\newcommand{\rnc}{\renewcommand}


\rnc{\baselinestretch}{1.24}    
\setlength{\jot}{6pt}       
\rnc{\arraystretch}{1.24}   

\makeatletter
\rnc{\theequation}{\thesection.\arabic{equation}}
\@addtoreset{equation}{section}
\makeatother



\nc{\be}{\begin{equation}}

\nc{\ee}{\end{equation}}

\nc{\bea}{\begin{eqnarray}}

\nc{\eea}{\end{eqnarray}}

\nc{\xx}{\nonumber\\}

\nc{\ct}{\cite}

\nc{\la}{\label}

\nc{\eq}[1]{(\ref{#1})}

\nc{\newcaption}[1]{\centerline{\parbox{6in}{\caption{#1}}}}

\nc{\fig}[3]{

\begin{figure}
\centerline{\epsfxsize=#1\epsfbox{#2.eps}}
\newcaption{#3. \label{#2}}
\end{figure}
}


\def\CA{{\cal A}}
\def\CC{{\cal C}}
\def\CD{{\cal D}}
\def\CE{{\cal E}}
\def\CF{{\cal F}}
\def\CG{{\cal G}}
\def\CH{{\cal H}}
\def\CK{{\cal K}}
\def\CL{{\cal L}}
\def\CM{{\cal M}}
\def\CN{{\cal N}}
\def\CO{{\cal O}}
\def\CP{{\cal P}}
\def\CR{{\cal R}}
\def\CS{{\cal S}}
\def\CU{{\cal U}}
\def\CW{{\cal W}}
\def\CY{{\cal Y}}


\def\IR{{\hbox{{\rm I}\kern-.2em\hbox{\rm R}}}}
\def\IB{{\hbox{{\rm I}\kern-.2em\hbox{\rm B}}}}
\def\IN{{\hbox{{\rm I}\kern-.2em\hbox{\rm N}}}}
\def\IC{\,\,{\hbox{{\rm I}\kern-.59em\hbox{\bf C}}}}
\def\IZ{{\hbox{{\rm Z}\kern-.4em\hbox{\rm Z}}}}
\def\IP{{\hbox{{\rm I}\kern-.2em\hbox{\rm P}}}}
\def\IH{{\hbox{{\rm I}\kern-.4em\hbox{\rm H}}}}
\def\ID{{\hbox{{\rm I}\kern-.2em\hbox{\rm D}}}}


\def\a{\alpha}
\def\b{\beta}
\def\ga{\gamma}
\def\d{\delta}
\def\ep{\epsilon}
\def\ph{\phi}
\def\k{\kappa}
\def\l{\lambda}
\def\m{\mu}
\def\n{\nu}
\def\th{\theta}
\def\rh{\rho}
\def\s{\sigma}
\def\t{\tau}
\def\w{\omega}
\def\G{\Gamma}


\def\half{\frac{1}{2}}
\def\dint#1#2{\int\limits_{#1}^{#2}}
\def\goto{\rightarrow}
\def\para{\parallel}
\def\brac#1{\langle #1 \rangle}
\def\grad{\nabla}
\def\curl{\nabla\times}
\def\div{\nabla\cdot}
\def\p{\partial}
\def\e{\epsilon_0}


\def\Tr{{\rm Tr}\,}
\def\det{{\rm det}}


\def\vare{\varepsilon}
\def\bz{\bar{z}}
\def\bw{\bar{w}}


\def\do{{\bf R}_{NC}^{4}}
\def\re{{\bf R}_{NC}^2}
\def\mi{{\bf R}_C^2}
\def\c{{\bf C}}
\def\z{{\bf Z}}

\begin{titlepage}

\hfill\parbox{5cm} {SOGANG-HEP 298/02}

\hfill\parbox{4cm} {{\tt hep-th/0206001}}

\vspace{25mm}

\begin{center}
{\Large \bf Propagators in Noncommutative Instantons}

\vspace{15mm}
Bum-Hoon Lee$^a$\footnote{bhl@ccs.sogang.ac.kr} and Hyun Seok
Yang$^b$\footnote{hsyang@phys.ntu.edu.tw}
\\[10mm]

$^a${\sl Department of Physics, Sogang University,
Seoul 121-742, Korea} \\
$^b${\sl Department of Physics, National Taiwan University,
Taipei 106, Taiwan, R.O.C.} \\
\end{center}

\thispagestyle{empty}

\vskip2cm


\centerline{\bf ABSTRACT} 
\vskip 4mm 
\noindent

We explicitly construct Green functions for a field in 
an arbitrary representation of gauge group 
propagating in noncommutative instanton backgrounds 
based on the ADHM construction. The propagators
for spinor and vector fields can be constructed in terms of 
those for the scalar field in noncommutative instanton background.
We show that the propagators in the adjoint representation 
are deformed by noncommutativity while 
those in the fundamental representation 
have exactly the same form as the commutative case. \\


\vspace{2cm}

\today

\end{titlepage}

\renewcommand{\thefootnote}{\arabic{footnote}}
\setcounter{footnote}{0}

\section{Introduction}

Instantons were found by Belavin, Polyakov, Schwartz and Tyupkin
(BPST) \ct{bpst} almost thirty years ago, as topologically nontrivial
solutions of the duality equations of the Euclidean Yang-Mills theory
with finite action. Immediately instantons were realized to describe
the tunnelling processes between different $\theta$-vacua in Minkowski
space and lead to the strong CP problem in QCD \ct{jare,cdg}.
(For the earlier development of instanton physics, see the collection
of papers \ct{shifman}.) The non-perturbative chiral anomaly in the
instanton background led to baryon number violation and a solution to
the $U(1)$ problem \ct{thooft1,thooft2}. These revealed that instantons can
have their relevance to phenomenological models like QCD and the
Standard model \ct{ss}.

Instanton solutions also appear as BPS states in string theory.
They are described by Dp-branes bound to D(p+4)-branes
\ct{witten,douglas}. Subsequently, in \ct{sv,cm}, low-energy
excitations of D-brane bound states were used to explain the
microscopic degrees of freedom of black-hole entropy, for which the
information on the instanton moduli space has a crucial role. In
addition the multi-instanton calculus was used for a non-perturbative test
of AdS/CFT correspondence \ct{bg,bgkr,dhkmv,bnv}, where
the relation between Yang--Mills instantons and D-instantons was
beautifully confirmed by the explicit form of the classical
D-instanton solution in $AdS_5 \times {\bf S}^5$ background
and its associated supermultiplet of zero modes.

Recently instanton solutions on noncommutative spaces have been turned out
to have more richer spectrums. While commutative instantons are always
BPS states, noncommutative instantons admit both BPS and non-BPS states.
Especially, instanton solutions can be found in $U(1)$ gauge theory
and the moduli space of non-BPS instantons is smooth,
small instanton singularities being resolved
by the noncommutativity \ct{ns,sw}. Remarkably, instanton solutions in
noncommutative gauge theory can also be studied by
Atiyah-Drinfeld-Hitchin-Manin (ADHM) equation \ct{adhm} slightly
modified by the noncommutativity \ct{ns}.
ADHM construction uses some quadratic matrix equations, hence
noncommutative objects in nature, to construct
(anti-)self-dual configurations of the gauge field.
Thus the noncommutativity of space doesn't make any serious obstacle
for the ADHM construction of noncommutative instantons and indeed it
turns out that it is a really powerful tool even for noncommutative
instantons. Recently much progress has been made in this direction
\ct{ns,sw,ly,bn,kf1,kko,ho,kly1,kf2,lty,kf3,
asch,corr,ckt,ham,kuro,kly2,lp,iks}.

In order to calculate instanton effects in quantum gauge theory,
it is important to know the Green function in instanton
backgrounds \ct{thooft2}. In this paper, based on the ADHM construction, 
we will construct the Green functions for a field in 
an arbitrary representation of gauge group 
propagating in noncommutative 
instanton backgrounds. Recently several papers
\ct{lty,dhk1,hkt,dhk2,klyi,holl,nekra,our,hiro} discussed the
instanton moduli space and 
the instanton calculus in noncommutative spaces.
This paper is organized as follows. In next section 
we review briefly the Weyl ordering prescription 
for operators and the Green function in noncommutative space, 
needed for later applications. In Section 3, 
we generalize the argument in \ct{bccl} to noncommutative space 
and show that the propagators
for spinor and vector fields can be constructed in terms of 
those for the scalar field in noncommutative instanton background.
In Section 4, we explicitly construct the scalar propagators in the
fundamental representation of $G$ and the tensor product 
$G_1 \times G_2$ \ct{cgt} 
where the adjoint representation is a special case.
We observe that the propagator in the adjoint representation or the
tensor product gauge group $G_1 \times G_2$ is deformed 
by noncommutativity while
that in the fundamental representation has exactly the same form
as the commutative case.
In Section 5 we speculatively discuss some important issues 
such as an infrared divergence in the vector propagator, 
the zero modes for the tensor product gauge group
and conformal property of instanton propagators.

\section{Green Function in Noncommutative Space}

In this section we review briefly the Weyl ordering prescription 
for operators and the Green function in noncommutative space \ct{chms,gn}, 
needed for later applications.

Here we will work in general in flat noncommutative Euclidean space 
${\bf R}^4$ represented by
\begin{equation}\label{nc-space}
  [\hat x^\mu, \hat x^\nu]=i\theta^{\mu\nu}
\end{equation}
where $\theta^{\mu\nu}=-\theta^{\nu\mu}$ and 
we use $\;{\hat{}}\;$ to indicate operators in $\CA_\theta$ 
for a moment. 
Since $\theta^{\mu\nu}$ is an anti-symmetric tensor, let's
decompose them into self-dual and anti-self-dual parts:
\begin{equation}\label{theta}
\theta_{\mu\nu} = \eta_{\mu\nu}^a \zeta^a + {\bar \eta}_{\mu\nu}^a
\chi^a.
\end{equation}
Since the self-duality condition is invariant under $SO(4)$ rotations
(or more generally $SL(4,{\bf R})$ transformations),
one can always make the matrix $\theta_{\mu\nu}$ to a standard
symplectic form by performing the $SO(4)$ transformation $R$: 
\be
\theta=R {\tilde \theta} R^T,
\ee
where we choose ${\tilde \theta}$ as
\be \la{theta}
{\tilde \theta}_{\mu\nu} = \left(
\begin{array}{cccc} 
0 & \theta_1 & 0 & 0 \\
-\theta_1 & 0 & 0 & 0 \\
0 & 0 & 0 & \theta_2 \\
0 & 0 & -\theta_2 & 0 
\end{array}
\right).
\ee
There are four important cases to consider:
\bea \la{c-space}
&& \bullet \; \theta_1=\theta_2=0: 
\quad \mbox{commutative ${\bf R}^4$},\\
\la{nc-sd}
&& \bullet \;  \theta_1=\theta_2={\zeta \over 4} : 
\quad \mbox{self-dual $\do$},\\
\la{nc-asd}
&& \bullet \;  \theta_1=-\theta_2={\zeta \over 4}: 
\quad \mbox{anti-self-dual $\do$},\\
\la{nc-22}
&& \bullet \;  \theta_1\theta_2=0 \; \mbox{but}\;  
\theta_1+\theta_2 = \frac{\zeta}{2}: 
\quad \re \times \mi.
\eea

By noncommutative space $\do$ one means the algebra $\CA_{\theta}$ 
generated by the $\hat x^\mu$ satisfying \eq{nc-space}.
The commutation relation \eq{nc-space} in the basis \eq{theta} 
is equivalent to that of the annihilation and creation operators 
for one dimensional or two-dimensional harmonic oscillator:
\be \la{harmonic}
 [a_a, a_b^\dagger]=\delta_{ab},
\ee
where $a=1,2$ for \eq{nc-sd} and \eq{nc-asd} 
and $a=1$ for \eq{nc-22}.
Explicitly, for self-dual and anti-self-dual $\do$ in \eq{nc-sd} 
and \eq{nc-asd}, 
\be \la{harmonic-4}
 a_a^\dagger=\sqrt{\frac{2}{\zeta}}(\hat x^{2a}+i\epsilon^{a-1} 
\hat x^{2a-1}), \quad
a_a =\sqrt{\frac{2}{\zeta}} (\hat x^{2a}-i \epsilon^{a-1} \hat x^{2a-1}),
\ee
where $a=1,2$ and $\epsilon = \theta_1/\theta_2$. 
So, for self-dual and anti-self-dual $\do$, the representation 
space $\CH$ of $\CA_\theta$ can be identified with 
the Fock space ${\cal F} = \sum_{(n_1,n_2) \in {\bf Z}_{\geq 0}^2 } 
{\bf C} | n_1 , n_2>$, where $n_1, n_2$ are occupation numbers 
in the harmonic oscillators.
Thus the noncommutative space $\do$ in the basis $\CF$ 
becomes two-dimensional integer lattice $\{(n_1, n_2) \in {\bf Z}_{\geq
0}^2\}$ and the integration on $\do$ can be defined 
by the sum over the lattice,
\be \la{int-4}
{\rm Tr}_\CH {\cal
O}(x)\equiv \Bigl(\frac{\zeta\pi}{2}\Bigr)^2 \sum_{(n_1,
n_2)}\langle n_1,n_2|{\cal O}(x)|n_1,n_2\rangle 
\ee 
for an operator ${\cal O}(x)$ in $\CA_\theta$.
While, for $\re \times \mi$ in \eq{nc-22},
\be \la{harmonic-22}
 a^\dagger=\frac{\hat x^{2a}+i \hat x^{2a-1}}{\sqrt{\zeta}}, \quad
a =\frac{\hat x^{2a}-i \hat x^{2a-1}}{\sqrt{\zeta}},
\ee
where $a=1$ for $\theta_1 \neq 0$ and $a=2$ for $\theta_2 \neq 0$. 
In this case, the representation space $\CH$ is given by
$\CF= \sum_{n \in {\bf Z}_{\geq 0}} {\bf C} | n\rangle$ 
and the integration for an operator ${\cal
O}(x)$ in $\CA_\theta$ with $\theta_1 \neq 0$, for example, 
can be replaced by
\begin{equation}
\la{int-22} \int d^4x {\cal O}(x) \rightarrow \zeta\pi \sum_{n \in
{\bf Z}_{\geq 0}} \int d^2x \langle n|{\cal O}(x)|n \rangle,
\end{equation}
where $d^2x=dx^3 dx^4$.

We introduce coherent states defined by
\be \la{coherent}
|\xi \rangle = e^{\xi a^\dagger} |0 \rangle, 
\qquad \langle \xi|=  \langle 0| e^{{\bar \xi} a}
\ee
where $|0 \rangle$ is a vacuum defined by $a|0 \rangle=0$. 
For notational simplicity, we only present the construction for the
algebra \eq{harmonic-22}, but the similar
construction can be given for \eq{harmonic-4}, 
for which $|\xi \rangle = e^{\xi^a a_a^\dagger} |0 \rangle$. 
The state $|\xi \rangle$ satisfies 
\be \la{coh-1}
a|\xi \rangle = \xi |\xi \rangle, \qquad  
\langle \xi|a^\dagger =  \langle \xi|{\bar \xi}
\ee
and 
\be \la{coh-2}
 \langle \eta| \xi \rangle = e^{{\bar \eta} \xi},
\quad \int \frac{d{\bar \xi}d\xi}{2\pi i} e^{-|\xi|^2}  
|\xi \rangle \langle \xi|=1.
\ee
Then we see that
\be \la{coh-3}
\langle \eta|e^{i(k_1\hat x^1+k_2 \hat x^2)}| \xi \rangle
=e^{-\frac{\zeta k^2}{8}} e^{i(k_1z^1+ k_2z^2)}  e^{{\bar \eta} \xi}
\ee
where $k^2=k_1^2 + k_2^2$ and $z^1=i\frac{\sqrt \zeta}{2}(\xi-{\bar \eta}),
\;z^2=\frac{\sqrt \zeta}{2}(\xi+{\bar \eta})$.

It is well-known that the Weyl or symmetric ordering
prescription provides the procedure that maps commutative smooth
functions onto operators acting on the Fock space $\CF$ \ct{gms}:
\be \la{weyl}
f(x) \longmapsto {\hat f}(\hat x) = \int \frac{d^4 k}{(2\pi)^4} 
f(k) e^{ik\cdot \hat x},
\ee
where 
\be \la{fourier}
f(k)= \int d^4 x f(x) e^{-ik\cdot x}.
\ee
Using the prescription \eq{weyl}, it is easy to show
that the operator multiplication in $\CA_{\theta}$ is isomorphic 
to the Moyal product of functions:
\be \la{fg}
{\rm If}\; f(x) \longmapsto  {\hat f}(\hat x) \; 
{\rm and}\; g(x) \longmapsto {\hat g}(\hat x), \;    
{\rm then}\; (f*g)(x) \longmapsto ({\hat f}{\hat g})(\hat x),
\ee
where the Moyal product is defined as
\be \la{moyal}
(f*g)(x)= e^{\frac{i}{2}\theta^{\mu\nu} \frac{\partial}{\partial x^\mu}
\frac{\partial}{\partial y^\nu}} f(x) g(y)|_{x=y}.
\ee

In order to discuss instanton propagators in the noncommutative 
space \eq{nc-space}, we first should know the free Green function 
${\hat G}^{(0)}(\hat x, \hat y)$ for the ordinary Laplacian \ct{chms,gn}: 
\be \la{free-green}
-{\hat \partial}_\mu {\hat \partial}_\mu 
{\hat G}^{(0)}(\hat x, \hat y)={\hat \delta}(\hat x - \hat y)
\ee
where the derivative for an operator $\hat f(\hat x)$ is defined as
\be \la{derivative}
{\hat \partial}_\mu \hat f(\hat x) = -i(\theta^{-1})_{\mu\nu}
[\hat x^\nu, \hat f(\hat x)].
\ee
In commutative ${\bf R}^4$, it is given by
\be \la{c-green}
G^{(0)}(x,y)=\frac{1}{4\pi^2(x-y)^2}.
\ee 
Here some comments should be made. In order to define the Green
function, we have introduced the tensor product 
$\CA^{1,2}_\theta = \CA^1_\theta \otimes \CA^2_\theta$ of 
two copies of the algebra $\CA_\theta$. We represent
$\CA^{1,2}_\theta$ as an algebra of operators on the tensor product 
$\CH^{1,2} = \CH^1 \otimes \CH^2$ of two Fock spaces. The functions 
${\hat G}^{(0)}(\hat x, \hat y),\; {\hat \delta}(\hat x - \hat y) \in 
\CA^{1,2}_\theta$ are operators acting on $\CH^{1,2}$. 
We identify $\hat x^\mu = \hat x^\mu \otimes 1$ and 
$\hat y^\mu = 1 \otimes \hat y^\mu$ in the tensor product. 
Thus in operator sense $[\hat x^\mu, \hat y^\nu]=0$. \footnote
{This is consistent with the fact that the Moyal bracket 
between two sets of independent variables vanishes, 
that is $x^\mu * y^\nu - y^\nu * x^\mu=0$ 
since $\frac{\p y^\nu}{\p x^\mu}=0$.} 
Therefore, if we introduce the ``center of mass coordinates'' $\hat R^\mu$ 
and the ``relative coordinates'' $\hat r^\mu$ defined by
\be \la{Rr}
\hat R^\mu = \frac{\hat x^\mu + \hat y^\mu}{2}, \qquad 
\hat r^\mu=\hat x^\mu-\hat y^\mu,
\ee
they satisfy the following commutation relations \ct{chms}
\be \la{com-Rr}
[\hat R^\mu, \hat R^\nu]=\frac{i}{2}\theta^{\mu\nu}, \quad
[\hat r^\mu, \hat r^\nu]=2i\theta^{\mu\nu}, 
\quad [\hat R^\mu, \hat r^\nu]=0.
\ee
The tensor product $\CA^{1,2}_\theta$ can be thus decomposed 
in the form
\be \la{tp-dr}
\CA^{1,2}_\theta \cong \CD \otimes \CR
\ee
where $\hat R^\mu$ acts on $\CD$ and $\hat r^\mu$ on $\CR$. Since the
noncommutative space \eq{nc-space} is homogeneous and 
so always respects a global translation symmetry, 
it is reasonable to require the translation
invariance for the Green function $\hat G^{(0)}(\hat x, \hat y)$. 
In other words the Green function depends only on $\hat r^\mu$. 
Using the Weyl prescription \eq{weyl}, we see that
\footnote{From now on, we will delete $\;{\hat{}}\;$ to indicate 
operators in $\CA_\theta$ for notational convenience 
as long as it doesn't cause any confusion.}  
\bea \la{fourier-green}
&& G^{(0)}(x-y) = \int \frac{d^4k}{(2\pi)^4} G^{(0)}(k) 
e^{ik\cdot (x-y)}, \\
&& \delta(x-y)= \int \frac{d^4k}{(2\pi)^4} 
e^{ik\cdot (x-y)}. \nonumber
\eea
Then the defining relation \eq{free-green} implies that 
$G^{(0)}(k)=\frac{1}{k^2}$.

To discuss more general Green functions, especially instanton
propagators, let's describe a formal procedure defining the Green
function. Let $\Delta$ be a linear operator 
on $\CA_\theta$ with a set
of eigenvectors $\phi_r(x) \in \CA_\theta$ 
and corresponding eigenvalues
$\lambda_r$:
\be \la{eigen}
\Delta \phi_r(x) = \lambda_r \phi_r(x),
\ee
where the parameter $r$ can be either continuous or discrete.
We shall assume the completeness of $\phi_r(x)$
\be \la{H-complete}
\Tr_\CH \phi_r(x)^\dagger \phi_s(x) = \delta_{rs}
\ee
in the Hilbert space $\CH_p$ of one-particle states to be 
\be \la{p-hilbert}
\CH_p = \{ \phi(x)= \sum_r a_r \phi_r(x) : \sum_r |a_r|^2 < \infty \}.
\ee
As usual the $a_r$ becomes operators (for example, 
creation or annihilation operators of a particle 
with quantum number $r$) when the field is quantized. 
The Green function is defined as the formal sum
\be \la{f-green}
G(x,y) = \sum_r \lambda_r^{-1} \phi_r(x) \phi_r (y)^\dagger.
\ee 
For the free Green function in \eq{free-green}, for example, 
$\phi_k(x) = e^{ik\cdot x}$ and $\lambda_k=k^2$ for the
Laplacian $\Delta = -\p_\mu \p_\mu$. In this case, the sum over $r$
should be the integration over momenta $k^\mu$ as in
\eq{fourier-green}.

\section{Instanton Propagators in Noncommutative Space}

In this section we will generalize the
argument in \ct{bccl} to noncommutative space to construct propagators
for spinor and vector fields in terms of those for the scalar field
in noncommutative instanton background. This generalization is
straightforward so one may regard it as a review of Sec. II and
Sec. III in \ct{bccl}. This result definitely generalizes that 
of free fields \ct{qft};
the Green functions for spinor and vector fields propagating in vacuum
are determined by the corresponding scalar propagator.

To consider spinor propagator, let's introduce quaternions defined by
\begin{equation}\label{quaternion}
  {\bf x}=x_\mu \sigma^\mu, \qquad {\bar{\bf x}}=x_\mu {\bar\sigma}^\mu,
\end{equation}
where $\sigma^\mu=(i \tau^a, 1)$ and ${\bar \sigma}^\mu=(-i
\tau^a, 1)=-\sigma^2 \sigma^{\mu T} \sigma^2$. 
The quaternion matrices $\sigma^\mu$ 
and ${\bar \sigma}^\mu$ have the basic properties
\bea \la{sigma}
&& \sigma^\mu {\bar \sigma}^\nu=\delta^{\mu\nu}+i\sigma^{\mu\nu},
\qquad \sigma^{\mu\nu}=\eta^a_{\mu\nu} \tau^a=*\sigma^{\mu\nu},\xx
&& {\bar \sigma}^\mu \sigma^\nu=\delta^{\mu\nu}+i{\bar
\sigma}^{\mu\nu}, \qquad {\bar \sigma}^{\mu\nu}={\bar
\eta}^a_{\mu\nu} \tau^a= -*{\bar \sigma}^{\mu\nu}
\eea
and
\bea \la{sigma-cp}
&& {\bar \sigma}^\mu_{\alpha\beta} \sigma^\mu_{\gamma\delta}=
\sigma^\mu_{\alpha\beta} {\bar \sigma}^\mu_{\gamma\delta}=
2\delta_{\alpha\delta}\delta_{\beta\gamma}, \xx
&& \sigma^\mu_{\alpha\beta} \sigma^\mu_{\gamma\delta}=
{\bar \sigma}^\mu_{\alpha\beta} {\bar \sigma}^\mu_{\gamma\delta}=
2\vare_{\alpha\gamma} \vare_{\beta\delta},
\eea
where $\alpha, \beta, \gamma, \delta=1,2$ are quaternionic indices.
The $\sigma^\mu$ and ${\bar \sigma}^\mu$ can be used to construct the
Euclidean Dirac matrices as
\bea \la{gamma}
&& \gamma^\mu =
\left(\begin{array}{cc}
0 & {\bar \sigma}^\mu \\
\sigma^\mu & 0 \end{array}\right), \qquad
\gamma_5 = \gamma_1 \gamma_2 \gamma_3 \gamma_4 =\left(\begin{array}{cc}
1 & 0 \\
0 & -1 \end{array}\right),\xx
&& \{ \gamma^\mu, \gamma^\nu \}=2 \delta^{\mu\nu}, \qquad
\gamma^{\mu\nu}=\frac{1}{2i}[\gamma^\mu, \gamma^\nu] = 
\left(\begin{array}{cc}
{\bar \sigma}^{\mu\nu} & 0 \\
0 & \sigma^{\mu\nu} \end{array}\right).
\eea
Thus \eq{sigma} and \eq{gamma} show that
\be \la{sd-gamma}
*\gamma^{\mu\nu} \frac{1 \pm \gamma_5}{2}=
\mp \gamma^{\mu\nu} \frac{1\pm \gamma_5}{2}.
\ee

We shall consider the propagator for spinor fields transforming in an 
arbitrary representation (fundamental, adjoint, etc.) of $U(N)$ gauge
group in the background of (anti-)self-dual instantons. The covariant
derivative $D_\mu$ is defined by
\be \la{coder}
D_\mu=\p_\mu + A_\mu
\ee
and the field strength $F_{\mu\nu}$ is given by
\bea \la{fsF}
F_{\mu\nu}&=&[D_\mu, D_\nu] \xx
&=&\partial_\mu A_\nu-\partial_\nu A_\mu +[A_\mu,A_\nu].
\eea
Since we are interested in spinor fields propagating in the
background of (anti-)self-dual instantons, we will assume that 
the field strength satisfies the (anti-)self-duality condition 
\begin{equation}\label{self-dual}
F_{\mu\nu}=\pm *F_{\mu\nu}=
\pm \half \vare_{\mu\nu\rho\sigma}F_{\rho\sigma}.
\end{equation}
Then we have
\be \la{gD2}
(\gamma \cdot D)^2 \frac{1 \pm \gamma_5}{2}=D^2 
\frac{1 \pm \gamma_5}{2}+\frac{i}{2} F^{\mu\nu}\gamma^{\mu\nu} 
\frac{1 \pm \gamma_5}{2}.
\ee
The equation \eq{sd-gamma} forces the second term of the right-hand side of
\eq{gD2} to vanish for spinors with positive (negative) chirality in
the self-dual (anti-self-dual) instanton background. In this case
there is no zero mode solution satisfying
\be \la{zero-spinor}
\gamma^\mu D_\mu \psi_{\pm}^{(0)}=0, \qquad \gamma_5 
\psi_\pm^{(0)} =\pm \psi_\pm^{(0)}.
\ee
However the second term in \eq{gD2} does not vanish 
for positive (negative) chirality spinor in the
anti-self-dual (self-dual) instantons. In this case
a finite number of zero modes satisfying \eq{zero-spinor} can be found. 
In the background of $k$ instantons in $U(N)$ gauge theory, 
the number of zero modes is $k$ in the fundamental representation and
$2Nk$ in the adjoint representation \ct{our}.

We will now consider a spinor field in the
background of $k$ anti-self-dual instantons. The self-dual case is
obtained simply by changing the sign of $\gamma_5$, $\gamma_5 \to
-\gamma_5$.
Let's introduce eigenfunctions $\psi_r(x)$ such that
\be \la{dirac-eigen}
\gamma^\mu D_\mu \psi_r(x)=\lambda_r \psi_r(x)
\ee
to define the spin-$\half$ Green function $S(x,y)$ which is described
by the formal expression
\be \la{green-spinor}
S(x,y) = {\sum_r}^\prime \lambda_r^{-1} \psi_r(x) 
\psi_r (y)^\dagger
\ee
where the prime means that the zero modes (states with
$\lambda_r=0$) are excluded from the sum.
It follows from \eq{green-spinor} that the spin-$\half$ propagator is
orthogonal to all the zero modes in \eq{zero-spinor}
\be \la{zero-ortho}
\Tr_\CH^x \Bigl( {\psi^{(0)}_n(x)}^\dagger S(x,y) \Bigr)=0.
\ee
Thus the spin-$\half$ propagator obeys the following equation
\be \la{dirac-green}
\gamma^\mu D_\mu S(x,y) = Q(x,y)
\ee
where
\be \la{q}
Q(x,y)=\delta(x,y)-\sum_n \psi^{(0)}_n(x)
{\psi^{(0)}_n(y)}^\dagger 
\ee
with the summation running over all the zero modes ($n=1,\cdots, k$ 
for spinors in the fundamental representation and $n=1, \cdots, 2Nk$
in the adjoint representation). The quantity $Q(x,y)$ represents the
projection operator, i.e. $\Tr_\CH^z \Bigl(Q(x,z)Q(z,y) \Bigr) 
=Q(x,y)$, into the subspace of all nonzero modes.

Using the same operator technique as in \ct{bccl}, 
the construction of $S(x,y)$ can be easily achieved. Let's introduce
an operator $S$ whose matrix representation with regard to 
position eigenstates in $\CH_p$ is $S(x,y)$,
\be \la{s}
\langle x| S |y \rangle = S(x,y).
\ee
Similarly, we write the corresponding spin-0 propagator $G(x,y)$,
which is defined by
\be \la{G}
-D_\mu D_\mu G(x,y)=\delta(x-y),
\ee
as the matrix element of an operator ${1 \over -D^2}$,
\be \la{g}
\langle x| \frac{1}{-D^2}|y \rangle = G(x,y).
\ee
We will show that the operator expression of the spin-$\half$
propagator is 
\be \la{S}
S = - \gamma \cdot D\frac{1}{-D^2}\frac{1 - \gamma_5}{2}  
- \frac{1}{-D^2} \gamma \cdot D \frac{1 + \gamma_5}{2}.
\ee
First note that 
\be \la{sq}
\gamma \cdot D S = Q,
\ee
where
\be \la{op-q}
Q=\frac{1 - \gamma_5}{2}- \gamma \cdot D  
\frac{1}{-D^2} \gamma \cdot D \frac{1 + \gamma_5}{2}.
\ee
The equation \eq{sq} implies that $Q$ contains no zero modes since
they are annihilated by $\gamma \cdot D$. On the other hand, 
we find that
\be \la{op-dq}
\gamma \cdot D Q = \gamma \cdot D
\ee
and
\be \la{op-qs}
QS=S.
\ee
It is easy to show that $Q^2=Q$. Therefore Eq.\eq{op-dq} shows that
$Q$ is the operator which projects into the subspace of all
nonzero modes and so $\langle x| Q |y \rangle$ is the function
defined in \eq{q}. Moreover, Eq.\eq{op-qs} implies that 
$S$ is orthogonal to all the zero modes. 
This ensures our claim in \eq{S}.

Let's consider Yang-Mills theory with gauge group $U(N)$ with action
\be \la{ym-action}
S=-\half \Tr_\CH {\rm tr}\Bigl( F_{\mu\nu} F_{\mu\nu}+
\frac{1}{\xi} (D_\mu A_\mu)^2\Bigr) 
\ee
and small fluctuations about a classical instanton
solution $A_\mu(x)$
\be \la{gauge-fluc}
A^\prime_\mu(x) = A_\mu(x) + \delta A_\mu(x).
\ee
If the action is expanded to second order in $\delta A_\mu$, 
one can find the following result 
\be \la{vary-action}
S[A^\prime_\mu] \approx S[A_\mu] -\Tr_\CH{\rm tr} \delta A_\mu
\Bigl(-D^2 \delta_{\mu\nu}-2 F_{\mu\nu}+(1-{1\over \xi})D_\mu D_\nu
\Bigr)\delta A_\nu.
\ee
In our previous paper \ct{our} 
we showed that in $k$ instanton background there
are $4Nk$ adjoint zero modes $\phi_\mu^{(n)},\; n=1,\cdots,4Nk$
satisfying $D_\mu \phi_\mu^{(n)}=0$ and
\be \la{zero-A}
(D^2 \delta_{\mu\nu}+2 F_{\mu\nu} \Bigr)\phi_\nu^{(n)} =0.
\ee
Thus to define spin-1 propagator we should project
out the zero modes just as the spin-$\half$ propagator \eq{dirac-green}.
According to the action \eq{vary-action}, the spin-1 propagator 
$G_{\mu\nu}(x,y)$ in the anti-self-dual $k$ instanton background is
defined by
\be \la{ym-green}
\Bigl(-D^2 \delta_{\mu\lambda}-2 F_{\mu\lambda}+(1-{1\over \xi})
D_\mu D_\lambda \Bigr)G_{\lambda\nu}(x,y)=Q_{\mu\nu}(x,y),
\ee
where
\be \la{ym-q}
Q_{\mu\nu}(x,y)=\delta_{\mu\nu}\delta(x-y)-\sum_n
\phi_\mu^{(n)}(x){\phi_\nu^{(n)}(y)}^\dagger.
\ee
The quantity $Q_{\mu\nu}(x,y)$ is the projection operator, 
i.e. $\Tr_\CH^z Q_{\mu\lambda}(x,z)Q_{\lambda\nu}(z,y)=
Q_{\mu\nu}(x,y)$, onto the space of the nonzero modes.

Using the operator formalism used in the spin-$\half$ propagator, 
one can show that the spin-1 propagator can also be constructed 
in terms of corresponding
scalar propagator. To proceed with the construction, define
\be \la{q-tensor}
q^{(\pm)}_{\mu\nu\lambda\kappa}=\delta_{\mu\nu}
\delta_{\lambda\kappa}+\eta^{(\pm)a}_{\mu\nu}
\eta^{(\pm)a}_{\lambda\kappa}
=\delta_{\mu\lambda}
\delta_{\nu\kappa}+\eta^{(\mp)a}_{\mu\lambda}
\eta^{(\mp)a}_{\nu\kappa}
\ee
where $\eta^{(+)a}_{\mu\nu} = \eta^a_{\mu\nu}$ and 
$\eta^{(-)a}_{\mu\nu} = {\bar \eta}^a_{\mu\nu}$ defined in \eq{sigma}.
The tensor
\be \la{proj-eta}
\eta^{(\pm)a}_{\mu\nu}\eta^{(\pm)a}_{\lambda\kappa}
=\frac{1}{4}(\delta_{\mu\lambda}\delta_{\nu\kappa}
-\delta_{\mu\kappa}\delta_{\nu\lambda} \pm 
\vare_{\mu\nu\lambda\kappa})
\ee
projects out the self-dual part or anti-self-dual part of an
antisymmetric tensor since $\eta^{(\pm)a}_{\mu\nu}
\eta^{(\mp)b}_{\mu\nu}=0$. Following \ct{bccl}, let's introduce a
bracket operation
\be \la{bracket}
\{ X \}^{(\pm)}_{\mu\nu} =q^{(\pm)}_{\mu\nu\lambda\kappa} 
D_\lambda X D_\kappa
\ee
for an arbitrary operator $X$. Then it is easy to see that
\be \la{d*b}
D_\nu \{ X \}^{(\pm)}_{\nu\mu}= D^2 X D_\mu + 
[F_{\mu\nu} \mp *F_{\mu\nu}, X D_\nu].
\ee
So if the field strength satisfies the self-duality condition
\eq{self-dual}, Eq.\eq{d*b} reduces to
\be \la{d-b1}
D_\nu \{ X \}^{(\pm)}_{\nu\mu}= D^2 X D_\mu.
\ee
Similarly,
\be \la{d-b2}
\{ X \}^{(\pm)}_{\mu\nu}D_\nu= D_\mu D^2 X.
\ee
Let's quote the following algebraic relation \ct{bccl}
\be \la{qq}
q^{(\pm)}_{\mu\lambda\nu\sigma}
q^{(\pm)}_{\lambda\kappa\tau\rho}
=\delta_{\sigma\tau}q^{(\pm)}_{\mu\kappa\nu\rho}
+r^{(\pm)}_{\mu\kappa\nu\rho\sigma\tau}
\ee
where
\be \la{tensor-r}
r^{(\pm)}_{\mu\kappa\nu\rho\sigma\tau}
=(\delta_{\mu\nu}\eta^{(\mp)c}_{\kappa\rho}
- \delta_{\kappa\rho}\eta^{(\mp)c}_{\mu\nu}
+\vare_{abc} \eta^{(\mp)a}_{\mu\nu}
\eta^{(\mp)b}_{\kappa\rho})\eta^{(\mp)c}_{\sigma\tau}
\ee
and thus it has the following duality property
\be \la{dual-r}
\half \vare_{\sigma\tau\upsilon\lambda}
r^{(\pm)}_{\mu\kappa\nu\rho\sigma\tau}
=\mp r^{(\pm)}_{\mu\kappa\nu\rho\upsilon\lambda}.
\ee
In the derivation of \eq{tensor-r}, we used
\be \la{eta^2}
\eta^{(\pm)a}_{\lambda\mu}\eta^{(\pm)b}_{\lambda\nu}
=\delta_{ab}\delta_{\mu\nu}+\vare_{abc}\eta^{(\pm)c}_{\mu\nu}.
\ee
Using these properties, the following bracket composition law 
can be derived
\be \la{composition}
\{ X \}^{(\pm)}_{\mu\lambda} \{ Y \}^{(\pm)}_{\lambda\nu} 
= \{ X D^2 Y \}^{(\pm)}_{\mu\nu}.
\ee

Now it is straightforward to see that $G_{\mu\nu}(x,y)$ has the
following formal operator expression
\be \la{op-G}
G_{\mu\nu} = - \Bigl\{ \Bigl(\frac{1}{D^2} \Bigr)^2 \Bigr\}_{\mu\nu} 
+ (1-\xi)D_\mu \Bigl(\frac{1}{D^2}\Bigr)^2 D_\nu.
\ee  
The reason is following. First note that
\be \la{op-ymgreen}
\Bigl(-D^2 \delta_{\mu\lambda}-2 F_{\mu\lambda}+(1-{1\over \xi})
D_\mu D_\lambda \Bigr)G_{\lambda\nu}=Q_{\mu\nu},
\ee
where
\be \la{op-ymq}
Q_{\mu\nu}= \Bigl\{ \frac{1}{D^2} \Bigr\}_{\mu\nu}.
\ee
We used \eq{d-b1} and \eq{d-b2} and the bracket composition
\eq{composition} in the derivation. Comparing with \eq{ym-green}, we
see that $Q_{\mu\nu}$ doesn't contain any zero modes. And, using the
composition law \eq{composition}, one can easily see 
that $Q_{\mu\nu}$ is a projection operator, i.e. 
$Q_{\mu\lambda}Q_{\lambda\nu}=Q_{\mu\nu}$. Indeed, $Q_{\mu\nu}$ is the
projection operator onto all the nonzero modes in \eq{ym-q} and thus
an operator realization of the projector $Q_{\mu\nu}(x,y)$ since it
satisfies the following equations
\bea \la{qg-g}
&& Q_{\mu\lambda}G_{\lambda\nu}=G_{\mu\nu},\xx
&&\Bigl(-D^2 \delta_{\mu\lambda}-2 F_{\mu\lambda}+(1-{1\over \xi})
D_\mu D_\lambda \Bigr)Q_{\lambda\nu}=
\Bigl(-D^2 \delta_{\mu\nu}-2 F_{\mu\nu}+(1-{1\over \xi})
D_\mu D_\nu \Bigr).
\eea
Thus we complete the proof of our claim in \eq{op-G}.

\section{Scalar Instanton Propagators}

In order to calculate instanton effects in quantum gauge theory,
it is important to know the Green function in instanton
backgrounds \ct{thooft2}. 
In previous section, following the same method in
\ct{bccl}, we showed that the propagators
for spinor and vector fields can be constructed 
in terms of those for the scalar field
in noncommutative instanton background.  
Thus, if we can find the scalar propagator $G(x,y)$ \eq{G} 
for fundamental representation or adjoint representation,
we know the spin-$\half$ propagator $S(x,y)$ for each
representation in terms of \eq{S} and
the spin-1 propagator $G_{\mu\nu}(x,y)$ in terms of \eq{op-G}. 
In commutative space, the scalar propagator in the
fundamental representation has a remarkably 
simple expression \cite{cfgt,csw} 
\begin{equation} \label{green}
G(x,y)=v(x)^\dagger G^{(0)}(x,y) v(y)
\end{equation}
where $v(x)$ is a function determining ADHM gauge field $A_\mu(x)$
by $A_\mu(x)=v(x)^\dagger \partial_\mu v(x)$. 
The scalar propagator in the adjoint representation has more
complicated expression of which we will present the explicit form.
We will first show that the scalar propagator in the noncommutative
instanton background has the exactly same form that \eq{green}.

To derive the above remarkable formulae, we need the following
basic properties in the ADHM construction \ct{adhm,cfgt,csw}. 
The gauge field with
instanton number $k$ for $U(N)$ gauge group is given in the form
\begin{equation}\label{A}
 A_\mu (x)=v(x)^{\dagger} \partial_\mu v(x)
\end{equation}
where $v(x)$ is $(N+2k)\times N$ matrix defined by the equations
\begin{eqnarray}
\label{normalization}
  && v(x)^\dagger v(x)=1,\\
\label{zero-mode} && v(x)^\dagger \Delta(x)=0.
\end{eqnarray}
In (\ref{zero-mode}), $\Delta(x)$ is a $(N+2k) \times 2k$ matrix,
linear in the position variable $x$, having the structure
\begin{equation}\label{Delta}
  \Delta (x)= \left \{ \begin{array}{l} a - b {\bf x},\qquad \mbox{
  self-dual instantons},\\
 a - b {\bar {\bf x}},\qquad \mbox{
  anti-self-dual instantons},
\end{array} \right.
\end{equation}
where $a, b$ are $(N+2k) \times 2k$ matrices. $v(x)$ can be
thought of as a map from an $N$-complex dimensional space $W$ to a
$N+2k$-complex dimensional space $V$. Thus $\Delta(x)$ must obey
the completeness relation
\begin{equation} \label{complete}
P(x) + \Delta(x) f(x) \Delta(x)^\dagger =1
\end{equation}
where $P(x)=v(x) v(x)^\dagger$. The matrices $a, b$ are
constrained to satisfy the conditions that $\Delta(x)^\dagger
\Delta(x)$ be invertible and that it commutes with the
quaternions. These conditions imply that $\Delta(x)^\dagger
\Delta(x)$ as a $2k \times 2k$ matrix has to be factorized as
follows
\begin{equation} \label{invertible}
\Delta(x)^\dagger \Delta(x) =f^{-1}(x) \otimes 1_2
\end{equation}
where $f^{-1}(x)$ is a $k \times k$ matrix and $1_2$ is a unit
matrix in quaternion space.

Given a pair of matrices $a, b$, (\ref{normalization}) and
(\ref{zero-mode}) define $A_\mu$ up to gauge equivalence.
Different pair of matrices $a, b$ may yield gauge equivalent
$A_\mu$ since (\ref{normalization}) and (\ref{zero-mode}) are
invariant under
\begin{equation}\label{ugl}
  a \rightarrow QaK, \quad b \rightarrow QbK,  \quad v \rightarrow Qv
\end{equation}
where $Q \in U(N+2k)$ and $K \in GL(k,\c)$. This freedom can be
used to put $a, b$ in the canonical forms
\be \la{ab}
a = \left(\begin{array}{c}
\lambda \\
\xi \end{array}\right), \qquad b=\left(\begin{array}{c}
0 \\
1_{2k} \end{array}\right),
\ee
where $\lambda$ is an $N \times 2k$ matrix and $\xi$ is a $2k \times
2k$ matrix. Here we decompose the matrix $\xi$ in the quaternionic
basis ${\bar \sigma}^\mu$ as a matter of convenience
\begin{equation}
\xi = \xi_\mu {\bar \sigma}^\mu,
\end{equation}
where $\xi_\mu$'s are $k \times k$ matrices. In the basis \eq{ab}, the
constraint \eq{invertible} boils down to
\bea \la{adhm1}
&& {\rm tr}_2 \tau^a a^\dagger a =
\left \{ \begin{array}{l} \theta^{\mu\nu}{\bar \eta}^a_{\mu\nu},
\qquad \mbox{self-dual instantons},\\
 \theta^{\mu\nu}\eta^a_{\mu\nu},\qquad \mbox{anti-self-dual instantons},
\end{array} \right. \\
\la{adhm2}
&& \xi_\mu^\dagger = \xi_\mu,
\eea
where ${\rm tr}_2$ is the trace over the quaternionic indices.

\subsection{Scalar Propagator in Fundamental Representation}

Now we will explain how to derive the formulae \eq{green}. 
First note that the
covariant derivative for a field $\Phi$ in the fundamental
representation of $U(N)$ has the simple expression
\begin{equation}\label{coder}
  D_\mu \Phi = (\partial_\mu + v^\dagger \partial_\mu v) \Phi =
  v^\dagger \partial_\mu(v \Phi).
\end{equation}
Using this relation, let's calculate $-D_\mu D_\mu G(x,y)$,
\begin{eqnarray}\label{cal1-green}
  &&\, - D_\mu D_\mu \Bigl(v(x)^\dagger G^{(0)}(x,y) v(y) \Bigr)
  \nonumber \\
  &=&- v(x)^\dagger \partial_\mu \Biggl(P(x)\partial_\mu \Bigl(
  P(x) G^{(0)}(x,y) v(y) \Bigr)\Biggr).
\end{eqnarray}
Note that $v(x)^\dagger P(x) =v(x)^\dagger$, so
 \begin{eqnarray}\label{cal2-green}
  &&-D_\mu D_\mu G(x,y)=- v(x)^\dagger 
\partial_\mu \Bigl(P(x) \partial_\mu P(x) \Bigr)
  G^{(0)}(x,y) v(y) \nonumber \\
&& - 2v(x)^\dagger \Bigl(\partial_\mu P(x) 
\partial_\mu G^{(0)}(x,y) \Bigr) v(y)
-v(x)^\dagger  \partial_\mu \partial_\mu G^{(0)}(x,y) v(y). 
\end{eqnarray}
Let's calculate the first term of the right-hand side in 
(\ref{cal2-green})
\begin{eqnarray} \label{cal3-green}
&& \, v(x)^\dagger 
\partial_\mu \Bigl( P(x) \partial_\mu P(x) \Bigr) \nonumber \\
&=&- v(x)^\dagger \partial_\mu \Bigl( P(x) \partial_\mu \Delta(x) 
f(x) \Delta(x)^\dagger \Bigr) \nonumber \\
&=& v(x)^\dagger  \partial_\mu \Delta(x) f(x) \Delta(x)^\dagger 
\partial_\mu \Delta(x) f(x) \Delta(x)^\dagger
-v(x)^\dagger  \partial_\mu \Delta(x) 
\partial_\mu f(x) \Delta(x)^\dagger \nonumber \\
&& -v(x)^\dagger \partial_\mu \Delta(x) f(x) 
\partial_\mu \Delta(x)^\dagger
\end{eqnarray}
where we used (\ref{zero-mode}) and (\ref{complete}). Also note that 
\begin{equation} \label{derf}
\partial_\mu f(x) =
-f(x) \Bigl( \partial_\mu \Delta(x)^\dagger \Delta(x) +
 \Delta(x)^\dagger \partial_\mu \Delta(x) \Bigr) f(x)
\end{equation} 
from (\ref{invertible}).

For explicit calculation, let's take the
anti-self-dual instanton with $\Delta (x)= a - b {\bar {\bf x}}$. 
The self-dual case can be similarly calculated. 
Using $\partial_\mu \Delta (x)= - b {\bar \sigma}_\mu$ 
and $\partial_\mu \Delta (x)^\dagger = - \sigma_\mu b^\dagger$ 
and the following formula 
\begin{equation} \label{formula}
{\bar \sigma}_\mu \Delta(x)^\dagger b 
 {\bar \sigma}_\mu = -2 b^\dagger \Delta(x), \qquad 
{\bar \sigma}_\mu f(x) \sigma_\mu = 4 f(x),
\end{equation}
we arrive at
\begin{equation} \label{cal5-green}
v(x)^\dagger 
\partial_\mu \Bigl( P(x) \partial_\mu P(x) \Bigr) =
-4 v(x)^\dagger b f(x) b^\dagger,
\end{equation}
where we used the fact that the function $f(x)$ commutes with 
${\bar \sigma}_\mu$ and $\sigma_\mu$.
Then our original equation (\ref{cal2-green}) reduces to
 \begin{equation}\label{cal6-green}
 -D_\mu D_\mu G(x,y)=2v(x)^\dagger 
\Bigl(2 b f(x) b^\dagger G^{(0)}(x,y) - 
\partial_\mu P(x) \partial_\mu G^{(0)}(x,y) \Bigr) v(y) + \delta(x-y) 
\end{equation}
where $-\partial_\mu \partial_\mu G^{(0)}(x,y) = \delta(x-y)$ is used. 
Note that the whole procedure above until (\ref{cal6-green}) is 
totally valid even for noncommutative space.

To arrive at our final destination \eq{G}, we must show that
\begin{equation} \label{cal7}
v(x)^\dagger \Bigl(2 b f(x) b^\dagger G^{(0)}(x,y) - 
\partial_\mu P(x) \partial_\mu G^{(0)}(x,y) \Bigr) v(y)= 0.
\end{equation}
First let's show (\ref{cal7}) in commutative ${\bf R}^4$, 
where we don't have to worry about the ordering problem, which will
also be helpful to find the noncommutative version.
If one notices that $\Delta(y)^\dagger v(y) =0$ and 
\begin{equation} \label{cal10}
\partial_\mu G^{(0)}(x,y)=- G^{(0)}(x,y)
\frac{2(x-y)_\mu}{(x-y)^2},
\end{equation}
the second term of \eq{cal7} can be written as
\begin{eqnarray} \label{cal11}
&& v(x)^\dagger \partial_\mu P(x) \partial_\mu G^{(0)}(x,y) v(y) 
\nonumber \\
&=&v(x)^\dagger b {\bar \sigma}_\mu f(x) \partial_\mu G^{(0)}(x,y)
\Bigl(\Delta(x)^\dagger-\Delta(y)^\dagger \Bigr) v(y) \nonumber \\
&=& 2 v(x)^\dagger b f(x) {\bar \sigma}_\mu \sigma_\nu b^\dagger
G^{(0)}(x,y)\frac{(x-y)_\mu}{(x-y)^2} (x-y)_\nu v(y).
\end{eqnarray}
Since ${\bar \sigma}_\mu \sigma_\nu=\delta_{\mu\nu} 
+ i {\bar \sigma}_{\mu\nu}$, 
(\ref{cal11}) exactly cancels the first term in (\ref{cal7}). 
Thus we proved \eq{green} in the commutative ${\bf R}^4$.

Before going on a tour to noncommutative space, let's explain
why we expect \eq{green} even for the noncommutative space. 
The relation (\ref{coder}) implies that if we define
\begin{equation} \label{hat} 
\hat{\Phi} = v \Phi \quad {\rm and} \quad 
D_\mu \hat{\Phi} = \widehat{D_\mu \Phi},
\end{equation}
we get
\begin{equation} \label{coder-pro}
D_\mu \hat{\Phi} = P \partial_\mu \hat{\Phi}.
\end{equation}
We may interpret this result as follows \ct{cfgt}. 
The matrix $v:W \to V$ maps 
$\Phi$ in the $N$-dimensional complex vector space $W$ with 
$\hat{\Phi}$ in an $N+2k$-dimensional complex vector space $V$ 
which lies in an $N$-dimensional subspace
of ${\bf  C}^{N+2k}$, i.e. the subspace
\begin{equation} \label{pmodule}
E_x = \{ \xi \;| \; P(x) \xi= \xi \}
\end{equation}
orthogonal to $\Delta(x)$ onto which $P$ is the projection operator. 
The collection of spaces $\{ E_x \}$ as $x$ varies over ${\bf R}^4$ or 
${\bf R}^4_{NC}$ forms a vector bundle and this vector bundle
precisely defines the ADHM gauge fields $A_\mu (x)$ through 
(\ref{coder-pro}). This is the statement 
of Serre-Swan theorem \ct{c*-algebra};
the vector bundle over a $C^*$-algebra ${\cal A}$ (which is
a complex Banach algebra with adjoint operation) is a
finitely generated projective module. (A module ${\cal E}$ is
projective if there exists another module ${\cal F}$ such that the
direct sum ${\cal E} \oplus {\cal F}$ is free, i.e. 
${\cal E} \oplus {\cal F} \cong {\cal A} \otimes \cdots \otimes {\cal
  A}$ as right ${\cal A}$-module.) 
Thus one can imagine the Green function $G(x,y)$ for the field $\Phi$ 
living in the ``nontrivial'' $N$-dimensional vector space $W$, 
which is defined as $G(x,y)=\langle \Phi(x), \Phi(y) \rangle$, 
is obtained by the map $v:W \to V$ from the Green function 
$G^{(0)}(x,y) \equiv \langle \hat{\Phi}(x), \hat{\Phi}(y) \rangle$ 
for the field $\hat{\Phi}$ living in the ``free'' $N+2k$-dimensional 
vector space $V$. This is precisely the content of (\ref{green}). 
Note, however, that this argument should
also be valid for a noncommutative space. This is the reason why we
expect the propagator (\ref{green}) even for noncommutative instanton
background.

To show (\ref{cal7}) in the noncommutative space \eq{nc-space}, 
let's chase the previous commutative calculation 
keeping in mind the ordering due to the noncommutativity. 
The second term in (\ref{cal7}) can be written as
\begin{eqnarray} \label{nccal1}
&& v(x)^\dagger \partial_\mu P(x) \partial_\mu G^{(0)}(x,y) v(y) \xx
&=& v(x)^\dagger b {\bar \sigma}_\mu f(x) 
\Delta(x)^\dagger \partial_\mu G^{(0)}(x,y) v(y) \nonumber \\
&=& v(x)^\dagger b f(x) {\bar \sigma}_\mu 
\partial_\mu G^{(0)}(x,y) \Bigl(\Delta(x)^\dagger- 
\Delta(y)^\dagger \Bigr) v(y)+ v(x)^\dagger b f(x) {\bar \sigma}_\mu 
[\Delta(x)^\dagger, \partial_\mu G^{(0)}(x,y)] v(y)\nonumber \\
&=&- v(x)^\dagger b f(x) {\bar \sigma}_\mu \sigma_\nu b^\dagger
\partial_\mu G^{(0)}(x,y) (x-y)_\nu v(y)
- v(x)^\dagger b f(x) {\bar \sigma}_\mu \sigma_\nu b^\dagger
[x_\nu, \partial_\mu G^{(0)}(x,y)] v(y)\nonumber \\
&=&- v(x)^\dagger b f(x) {\bar \sigma}_\mu \sigma_\nu b^\dagger
(x-y)_\nu \partial_\mu G^{(0)}(x,y) v(y)
- v(x)^\dagger b f(x) {\bar \sigma}_\mu \sigma_\nu b^\dagger
[y_\nu, \partial_\mu G^{(0)}(x,y)] v(y)\nonumber \\
&=&- \half v(x)^\dagger b f(x) {\bar \sigma}_\mu \sigma_\nu b^\dagger
\Bigl( \partial_\mu G^{(0)}(x,y) (x-y)_\nu+
(x-y)_\nu \partial_\mu G^{(0)}(x,y) \Bigr) v(y) \nonumber \\
&&- \half v(x)^\dagger b f(x) {\bar \sigma}_\mu \sigma_\nu b^\dagger
[(x+y)_\nu, \partial_\mu G^{(0)}(x,y)] v(y) \nonumber \\
&=&- \half v(x)^\dagger b f(x) {\bar \sigma}_\mu \sigma_\nu b^\dagger
\Bigl( \partial_\mu G^{(0)}(x,y) (x-y)_\nu +
(x-y)_\nu \partial_\mu G^{(0)}(x,y) \Bigr) v(y).
\end{eqnarray} 
In the last step, we used the fact that the function 
$\partial_\mu G^{(0)}(x,y)$ depends only on the combination $(x-y)$
because of translation invariance and \eq{com-Rr}.

In order to calculate the right-hand side of \eq{nccal1}, we will use
the Weyl symmetric prescription \eq{weyl}:
\bea \la{nccal4}
&&  \half {\bar \sigma}_\mu \sigma_\nu 
\Bigl( \partial_\mu G^{(0)}(x,y) (x-y)_\nu +
(x-y)_\nu \partial_\mu G^{(0)}(x,y) \Bigr) \xx 
&=& \int \frac{d^4k}{(2\pi)^4} {\bar \sigma}_\mu \sigma_\nu 
\frac{k_\mu}{k^2}\frac{\partial}
{\partial k^\nu} e^{ik\cdot (x-y)} \xx
&=& \int \frac{d^4k}{(2\pi)^4} 
{\bar \sigma}_\mu \sigma_\nu \frac{\partial}
{\partial k^\nu} \Bigl(\frac{k_\mu}{k^2} e^{ik\cdot (x-y)} \Bigr)
- \int \frac{d^4k}{(2\pi)^4}  {\bar \sigma}_\mu \sigma_\nu \frac{1}{k^2} 
\Bigl(\delta_{\mu\nu}-2 \frac{k_\mu k_\nu}{k^2}\Bigr) 
e^{ik\cdot (x-y)} \xx
&=& \int \frac{d^4k}{(2\pi)^4} {\bar \sigma}_\mu \sigma_\nu 
\frac{\partial} {\partial k^\nu} \Bigl(\frac{k_\mu}{k^2} 
e^{ik\cdot (x-y)} \Bigr) -2 G^{(0)}(x,y).
\eea
Thus if we can show that the total divergence 
$\int \frac{d^4k}{(2\pi)^4} {\bar \sigma}_\mu \sigma_\nu 
\frac{\partial} {\partial k^\nu} \Bigl(\frac{k_\mu}{k^2} 
e^{ik\cdot (x-y)} \Bigr) \equiv K(x-y)$ vanishes
on ${\bf S}^3$ in the large $k$ limit, we can finally achieve our
goal \eq{cal7} in the noncommutative space. 
It is easy to show directly in the basis of tensor product 
$|\xi_1, \xi_2 \rangle = |\xi_1 \rangle \otimes |\xi_2 \rangle $ of 
coherent states such as \eq{coherent}, using \eq{coh-3}, that 
the function $\langle \xi_1, \xi_2| K(x-y)|\xi_1, \xi_2 \rangle$ 
vanishes for any $|\xi_1, \xi_2 \rangle$.
This means that the operator function $K(x-y)$ 
should vanish even in the noncommutative space.

\subsection{Scalar Propagator in Adjoint Representation}

Next let's consider the scalar propagator in the adjoint
representation.
If $q$ denotes the fundamental representation of $U(N)$, the adjoint
representation can be obtained by tensor product $q \otimes {\bar q}$,
for which
\be \la{der-2f}
D_\mu =\p_\mu + A_\mu \otimes 1 + 1 \otimes {\bar A}_\mu.
\ee
In other word, we regard a field in the adjoint representation as a
two index object, one index transforming according to the fundamental
representation and the other its complex conjugate. Motivated by this
fact and to follow the method in \ct{cgt}, we treat this problem in a
more general context. Consider the direct product $G_1 \times
G_2$ and suppose we have instanton solutions
\be \la{2A}
A_\mu^1(x)=v_1(x)^\dagger \p_\mu v_1(x), \qquad  
A_\mu^2(x)=v_2(x)^\dagger \p_\mu v_2(x),
\ee
described in the ADHM way for each gauge group. Also consider a field
transforming under the fundamental representation of each and thus its
covariant derivative is defined by
\be \la{cod-2f}
D_\mu =\p_\mu + A^1_\mu \otimes 1 + 1 \otimes A^2_\mu.
\ee
The adjoint representation of $U(N)$ would be obtained by taking
$G_1=G_2=U(N)$ and $A_\mu^1=A_\mu^2=A_\mu$.

The Green function for a tensor product should be obtained by solving 
\eq{G} with $D_\mu$ defined by \eq{cod-2f}. 
Thus we will also consider the tensor product 
\be \la{v-tensor}
v(x)=v_1(x) \otimes v_2(x)
\ee
of two independent fields $v_1(x),\;v_2(x)$ in the fundamental
representation of $G_1$ and $G_2$ respectively. 
To preserve the group structure for $G_1 \times G_2$, i.e. 
$(g_1,g_2)(h_1,h_2)=(g_1h_1, g_2h_2) \in G_1 \times G_2$ for all $g_1,
h_1 \in G_1$ and $g_2, h_2 \in G_2$, 
we define a (unique) multiplication between 
elements of $G_1 \times G_2$ such that
\be \la{multi-law}
\Bigr(\phi_1(x) \otimes \phi_2(x)\Bigr)
\Bigr(\chi_1(x) \otimes \chi_2(x)\Bigr)
=\Bigr(\phi_1(x)\chi_1(x)\Bigr) \otimes
\Bigr(\phi_2(x) \chi_2(x)\Bigr)
\ee
for all $\phi_1(x),\; \chi_1(x) \in G_1$ and $\phi_2(x),\; \chi_2(x)
\in G_2$. This multiplication law will be crucial for our calculation
of adjoint Green function. 
The commutative Green function $G(x,y)$ satisfying
\eq{G} for $G_1 \times G_2$ was previously constructed in \ct{cgt}
and to be of the form
\be \la{ad-green}
G(x,y)=[v_1(x) \otimes v_2(x)]^\dagger G^{(0)}(x,y)[v_1(y) \otimes v_2(y)]
+\frac{1}{4\pi^2}C(x,y).
\ee
We will specify the explicit form of $C(x,y)$ later.

We will try the same expression as \eq{ad-green} to construct
noncommutative instanton propagator. 
To calculate \eq{G} for the ansatz \eq{ad-green},
first note that, using the multiplication law \eq{multi-law},
\bea \la{cod1-tensor}
D_\mu v(x)^\dagger &=& 
v_1(x)^\dagger \p_\mu P_1(x) \otimes v_2(x)^\dagger 
+ v_1(x)^\dagger \otimes v_2(x)^\dagger \p_\mu P_2(x),\\
D_\mu D_\mu v(x)^\dagger
&=& v_1(x)^\dagger \p_\mu \Bigl(v_1(x)^\dagger \p_\mu P_1(x)
\Bigr) \otimes v_2(x)^\dagger + v_1(x)^\dagger \otimes v_2(x)^\dagger 
\p_\mu \Bigl(v_2(x)^\dagger \p_\mu P_2(x)\Bigr) \xx
&& +2 v_1(x)^\dagger \p_\mu P_1(x)
 \otimes v_2(x)^\dagger \p_\mu P_2(x),
\eea
where we defined 
\be \la{2p}
P_a(x)=v_a(x)v_a(x)^\dagger, \qquad a=1,2.
\ee
If we further define 
\be \la{p-tensor}
P(x) = P_1(x) \otimes P_2(x),
\ee 
the above covariant derivatives can be rewritten as, using \eq{multi-law}
again,
\bea \la{cod2-tensor}
D_\mu v(x)^\dagger &=& 
v(x)^\dagger \p_\mu P(x),\\
D_\mu D_\mu v(x)^\dagger
&=& v(x)^\dagger \p_\mu \Bigl(P(x) \p_\mu P(x)\Bigr),
\eea
where we used $v_1^\dagger(x) P_1(x)=v^\dagger_1(x)$ and 
$v_2^\dagger(x) P_2(x)=v^\dagger_2(x)$. Thus we can proceed with 
the calculation of $- D_\mu D_\mu \Bigl(v(x)^\dagger G^{(0)}(x,y) v(y)
\Bigr)$ in the same way as Sec.3.1. Consequently we get
\bea \la{ad-first}
 &&- D_\mu D_\mu \Bigl(v(x)^\dagger G^{(0)}(x,y) v(y) \Bigr)
= \delta(x-y) \xx 
 &&-2 \Bigl(v_1(x)^\dagger  \p_\mu P_1(x) \otimes 
v_2(x)^\dagger  \p_\mu P_2(x)\Bigr)G^{(0)}(x,y) 
\Bigl(v_1(y) \otimes v_2(y)\Bigr).
\eea

Let's calculate the second term in \eq{ad-first}. 
For explicit calculation, let's take the
anti-self-dual instanton with $\Delta_a (x)= a_a - b_a {\bar {\bf x}}$. 
The self-dual case can be similarly done. 
Note that, according to ADHM construction, 
$v_a(x)^\dagger  \p_\mu P_a(x)=v_a(x)^\dagger b_a {\bar \sigma}_\mu 
f_a(x) \Delta_a(x)^\dagger$. Again the multiplication
law \eq{multi-law} will have a crucial role in the calculation below,
\bea \la{ad-cal1}
&&-2 \Bigl(v_1(x)^\dagger  \p_\mu P_1(x) \otimes 
v_2(x)^\dagger  \p_\mu P_2(x)\Bigr)G^{(0)}(x,y) v(y) \xx
&=&-2 \Bigl(v_1(x)^\dagger b_1 {\bar \sigma}_\mu 
f_1(x) \Delta_1(x)^\dagger \otimes 
v_2(x)^\dagger  \p_\mu P_2(x)\Bigr)G^{(0)}(x,y) v(y) \xx
&=&-2 \Bigl(v_1(x)^\dagger b_1 {\bar \sigma}_\mu 
f_1(x) \otimes v_2(x)^\dagger  \p_\mu P_2(x)\Bigr) \times \xx
 && \; \Bigl( [ \Delta_1(x)^\dagger \otimes 1, G^{(0)}(x,y)]
+G^{(0)}(x,y) \Bigl( \Delta_1(x)^\dagger 
-\Delta_1(y)^\dagger \Bigr) \otimes 1 \Bigr) v(y) \xx
&=&-2 \Bigl(v_1(x)^\dagger b_1 {\bar \sigma}_\mu 
f_1(x) \otimes v_2(x)^\dagger  \p_\mu P_2(x)\Bigr) \times \xx
 && \; \Bigl( [ \sigma_\nu b_1^\dagger y_\nu \otimes 1, G^{(0)}(x,y)]
+\Bigl(\sigma_\nu b_1^\dagger (x-y)_\nu 
\otimes 1 \Bigr)G^{(0)}(x,y) \Bigr) v(y) \xx
&=&- \Bigl(v_1(x)^\dagger b_1 {\bar \sigma}_\mu 
f_1(x) \otimes v_2(x)^\dagger  \p_\mu P_2(x)\Bigr) \times \xx
 && \; \Bigl( G^{(0)}(x,y)\Bigl(\sigma_\nu b_1^\dagger (x-y)_\nu 
\otimes 1 \Bigr) +\Bigl(\sigma_\nu b_1^\dagger (x-y)_\nu 
\otimes 1 \Bigr)G^{(0)}(x,y) \Bigr) v(y).
\eea
In the last step, we used the fact that the function 
$\partial_\mu G^{(0)}(x,y)$ depends only on the combination $(x-y)$
and \eq{com-Rr}. We can repeat the same procedure in \eq{ad-cal1} for
the term $v_2(x)^\dagger  \p_\mu P_2(x)$. Then we get
\bea \la{ad-cal2}
&&-2 \Bigl(v_1(x)^\dagger  \p_\mu P_1(x) \otimes 
v_2(x)^\dagger  \p_\mu P_2(x)\Bigr)G^{(0)}(x,y) v(y) \xx
&=&- \half \Bigl(v_1(x)^\dagger b_1 {\bar \sigma}_\mu 
f_1(x) \otimes v_2(x)^\dagger b_2 {\bar \sigma}_\mu 
f_2(x)\Bigr) \times \xx
 && \; \Bigl( F^{(0)}(x,y)\Bigl(1 \otimes \sigma_\lambda b_2^\dagger 
(x-y)_\lambda \Bigr) +\Bigl(1 \otimes \sigma_\lambda b_2^\dagger 
(x-y)_\lambda \Bigr)F^{(0)}(x,y) \Bigr) v(y),
\eea
where $F^{(0)}(x,y)=
 G^{(0)}(x,y) \Bigl(\sigma_\nu b_1^\dagger (x-y)_\nu 
\otimes 1 \Bigr) +\Bigl( \sigma_\nu b_1^\dagger (x-y)_\nu 
\otimes 1 \Bigr)G^{(0)}(x,y)$.

In order to calculate the right-hand side of \eq{ad-cal2}, 
we will again use the Weyl symmetric prescription \eq{weyl}.
First note that
\be \la{ad-weyl1}
F^{(0)}(x,y) = -2i ( \sigma_\nu b_1^\dagger \otimes 1) 
\int \frac{d^4k}{(2\pi)^4}  
\frac{1}{k^2}\frac{\partial}
{\partial k^\nu} e^{ik\cdot (x-y)}
\ee
and
\bea \la{as-weyl2}
&& F^{(0)}(x,y)\Bigl(1 \otimes \sigma_\lambda 
b_2^\dagger (x-y)_\lambda \Bigr) +\Bigl(1 \otimes \sigma_\lambda
b_2^\dagger (x-y)_\lambda \Bigr)F^{(0)}(x,y) \xx
&=& -4 ( \sigma_\nu b_1^\dagger \otimes  \sigma_\lambda b_2^\dagger) 
\int \frac{d^4k}{(2\pi)^4}  
\frac{1}{k^2}\frac{\partial^2}
{\partial k^\nu \partial k^\lambda} e^{ik\cdot (x-y)} \xx
&\equiv& -4 ( \sigma_\nu b_1^\dagger \otimes  \sigma_\lambda b_2^\dagger) 
H_{\nu\lambda} (x,y).
\eea
Then we arrive at
\bea \la{ad-cal3}
&&-2 \Bigl(v_1(x)^\dagger  \p_\mu P_1(x) \otimes 
v_2(x)^\dagger  \p_\mu P_2(x)\Bigr)G^{(0)}(x,y) v(y) \xx
&=& 2 \Bigl(v_1(x)^\dagger b_1 {\bar \sigma}_\mu \sigma_\nu 
f_1(x) b_1^\dagger \otimes v_2(x)^\dagger b_2 {\bar \sigma}_\mu 
\sigma_\lambda f_2(x) b_2^\dagger \Bigr) H_{\nu\lambda} (x,y) v(y), \xx
&=& 2 \Bigl(v_1(x)^\dagger b_1 {\bar \sigma}_\mu 
f_1(x) b_1^\dagger \otimes v_2(x)^\dagger b_2 {\bar \sigma}_\mu 
f_2(x) b_2^\dagger \Bigr) H_{\nu\nu} (x,y) v(y).
\eea
To derive the last result in \eq{ad-cal3}, we used \eq{sigma}, 
\eq{eta^2} and the fact that $H_{\nu\lambda} (x,y)= 
H_{\lambda\nu} (x,y)$.

In order to calculate $H_{\nu\nu}(x,y)$, we introduce an infrared cutoff 
$\epsilon$, i.e. $\frac{1}{k^2} \to \frac{1}{k^2 + \epsilon}$. After
the integral, we will take the limit $\epsilon \to 0$:
\bea \la{ad-int1}
 H_{\nu\nu} (x,y)&=& \lim_{\epsilon \to 0} 
\int \frac{d^4k}{(2\pi)^4} \frac{\p}{\p k_\nu}\Bigl( 
\frac{1}{k^2+\epsilon} \frac{\partial}
{\partial k^\nu} e^{ik\cdot (x-y)} + 
\frac{2k^\nu}{(k^2+\epsilon)^2} e^{ik\cdot (x-y)} \Bigr) \xx
&& -\lim_{\epsilon \to  0} \int \frac{d^4k}{(2\pi)^4} 
\frac{8 \epsilon}{(k^2+\epsilon)^3}
e^{ik\cdot (x-y)}.
\eea
One can check that 
\be \la{delta-fn}
\lim_{\epsilon \to 0}\frac{2}{\pi^2}\frac{\epsilon}{(k^2+\epsilon)^3} 
=\delta(k).
\ee
Using the coherent state basis 
$|\xi_1, \xi_2 \rangle = |\xi_1 \rangle \otimes |\xi_2 \rangle $ 
as in \eq{nccal4}, it is easy to see 
that the total divergence in \eq{ad-int1} vanishes
on ${\bf S}^3$ in the large $k$ limit. So we have
\be \la{h}
 H_{\nu\nu} (x,y)=-\frac{1}{4\pi^2}.
\ee
Finally we get
\bea \la{first-final}
&&- D_\mu D_\mu \Bigl(v(x)^\dagger G^{(0)}(x,y) v(y) \Bigr) \xx
&& = \delta(x-y)-\frac{1}{2\pi^2}  
\Bigl(v_1(x)^\dagger b_1 {\bar \sigma}_\mu 
f_1(x) b_1^\dagger \otimes v_2(x)^\dagger b_2 {\bar \sigma}_\mu 
f_2(x) b_2^\dagger \Bigr) v(y).
\eea
Note that the above expression is exactly the same form
as the commutative case.

Therefore, in order to get the answer \eq{G} for the spin-1 
propagator \eq{ad-green}, we must show that 
\bea \la{green-c}
&&-D_\mu D_\mu C(x,y)= \xx
&&  4 \Biggl( \Bigl(v_1(x)^\dagger b_1 f_1(x) \Bigr)_\alpha 
\otimes \Bigl(v_2(x)^\dagger b_2
f_2(x) \Bigr)_\beta \Biggr)
\Biggl( \Bigl(b_1^\dagger v_1(y) \Bigr)_\gamma \otimes 
\Bigl( b_2^\dagger v(y) \Bigr)_\delta \Biggr) 
\vare_{\alpha\beta} \vare_{\gamma\delta},
\eea
where we used \eq{sigma-cp} and \eq{multi-law}. Since the right-hand
side of \eq{green-c} has exactly the same form as the commutative one,
we will take the same ansatz for $C(x,y)$ with commutative case \ct{cgt}:
\be \la{c-com}
C_{us,vt}(x,y)=M_{ij,lm} \Bigl(v_1(x)^\dagger b_1 \Bigr)_{u,i\alpha} 
\Bigl( v_2(x)^\dagger b_2 \Bigr)_{s,j\beta}
\Bigl( b_1^\dagger v_1(y) \Bigr)_{l\gamma,v}
\Bigl( b_2^\dagger v_2(y) \Bigr)_{m\delta,t}
\vare_{\alpha\beta} \vare_{\gamma\delta}, 
\ee
or in the tensor notation
\be \la{c-ten}
C(x,y)= \Biggl( \Bigl(v_1(x)^\dagger b_1 \Bigr)_{\alpha} 
\otimes \Bigl( v_2(x)^\dagger b_2 \Bigr)_{\beta} \Biggr)
M \Biggl( \Bigl( b_1^\dagger v_1(y) \Bigr)_{\gamma}
\otimes \Bigl( b_2^\dagger v_2(y) \Bigr)_{\delta} \Biggr)
\vare_{\alpha\beta} \vare_{\gamma\delta},
\ee
where $M_{ij,lm}$ is a constant matrix to be determined later and 
$u,v=1,\cdots, N_1={\rm dim}G_1,\; s,t=1,\cdots, N_2={\rm dim}G_2$ are 
group indices and $i,l=1,\cdots,k_1,\; j,m=1,\cdots,k_2$ 
are instanton number indices.
Using the formula (see Eqs. \eq{coder} and \eq{cal5-green})
\bea \la{adhm-formula}
&& D_\mu v_a(x)^\dagger = v_a(x)^\dagger b_a 
{\bar \sigma}_\mu f_a(x) \Delta_a(x)^\dagger, \xx
&& D_\mu D_\mu v_a(x)^\dagger = -4 v_a(x)^\dagger b_a 
f_a(x) b_a^\dagger
\eea
and the multiplication \eq{multi-law},
it is straightforward to show that
\bea \la{c-cal1}
&& -D_\mu D_\mu C(x,y) = 
4 \Biggl( \Bigl(v_1(x)^\dagger b_1 f_1(x) \Bigr)_{\alpha} 
\otimes \Bigl( v_2(x)^\dagger b_2 f_2(x) \Bigr)_{\beta} \Biggr)  \xx
&& \Biggl( b_1^\dagger b_1 \otimes f^{-1}_2(x) 
+ f^{-1}_1(x) \otimes  b_2^\dagger b_2 -
\Bigl(\Delta_1(x)^\dagger b_1 \Bigr)_{\eta\xi} \otimes
\Bigl( \Delta_2(x)^\dagger b_2 \Bigr)_{\chi \zeta}
\vare_{\eta\chi} \vare_{\xi\zeta} \Biggr) M  \xx
&& \Biggl( \Bigl( b_1^\dagger v_1(y) \Bigr)_{\gamma}
\otimes \Bigl( b_2^\dagger v_2(y) \Bigr)_{\delta} \Biggr)
\vare_{\alpha\beta} \vare_{\gamma\delta},
\eea
where we used \eq{sigma-cp} at the final stage. Thus if the matrix $M$
satisfies 
\be \la{c-cal2}
 \Biggl( b_1^\dagger b_1 \otimes f^{-1}_2(x) 
+ f^{-1}_1(x) \otimes  b_2^\dagger b_2 -
\Bigl(\Delta_1(x)^\dagger b_1 \Bigr)_{\eta\xi} \otimes
\Bigl( \Delta_2(x)^\dagger b_2 \Bigr)_{\chi \zeta}
\vare_{\eta\chi} \vare_{\xi\zeta} \Biggr) = M^{-1},
\ee
we finally prove Eq.\eq{green-c} and so \eq{G} for the tensor product
$G_1 \times G_2$ where the adjoint representation is a special case.
However, to achieve this final goal, we should go further since
Eq.\eq{c-cal2} appears to state that the constant matrix $M$ is the
inverse of an $x$-dependent matrix. In commutative space, as a result
of conformal invariance of the matrix $M$ \ct{cgt}, 
the $x$-dependent parts of
the left-hand side in \eq{c-cal2} are completely canceled. We will
show that it is also the case even for noncommutative space, 
but the matrix $M$ is slightly modified by the noncommutativity.

First note that, in the canonical basis \eq{ab},
\be \la{c-cal3} 
f_a^{-1}(x)=\Delta_a(x)^\dagger \Delta_a(x) 
= a_a^\dagger a_a -\xi^\mu_a x_\mu 
+ x^2 -\half \sigma_{\mu\nu} \theta^{\mu\nu}
\ee
and
\bea \la{c-cal4}
\Bigl(\Delta_1(x)^\dagger b_1 \Bigr)_{\eta\xi} \otimes
\Bigl( \Delta_2(x)^\dagger b_2 \Bigr)_{\chi \zeta}
\vare_{\eta\chi} \vare_{\xi\zeta} &=&
(a_1^\dagger b_1)_{\eta\xi} \otimes
(a_2^\dagger b_2)_{\chi \zeta} \vare_{\eta\chi} \vare_{\xi\zeta} \xx
&&-2 x_\mu(\xi^\mu_1 \otimes 1)- 2 x_\mu (1 \otimes \xi^\mu_2) + 2 x^2.
\eea
Using these results, it is easy to see that the $x$-dependent 
parts of the left-hand side in \eq{c-cal2} are completely 
canceled and the matrix $M$ is defined by
\be \la{matrix-M}
 \Biggl( b_1^\dagger b_1 \otimes  (a_2^\dagger a_2- 
\half \sigma_{\mu\nu} \theta^{\mu\nu})
+ ( a_1^\dagger a_1 -\half \sigma_{\mu\nu} \theta^{\mu\nu})
\otimes  b_2^\dagger b_2 -(a_1^\dagger b_1)_{\eta\xi} \otimes
(a_2^\dagger b_2)_{\chi \zeta} 
\vare_{\eta\chi} \vare_{\xi\zeta} \Biggr) = M^{-1}.
\ee
Note that $a_a^\dagger a_a -\half \sigma_{\mu\nu} \theta^{\mu\nu}$ is
proportional to the identity matrix in quaternionic space while 
$a_a^\dagger a_a$ is not, as seen from \eq{c-cal3}. We see that the
matrix $M$ is deformed by the noncommutativity, but only for non-BPS
instanton background, that is anti-self-dual (self-dual) instantons 
in self-dual (anti-self-dual) $\do$ and all instantons for \eq{nc-22}, 
since $\eta^a_{\mu\nu} {\bar \eta}^b_{\mu\nu}=0$. 
If $\theta^{\mu\nu}=0$, of course, we recover the result in the
commutative space.

In order to construct the propagator for $q \otimes {\bar q}$ 
in \eq{der-2f}, that is, the adjoint representation of $U(N)$, 
we take \ct{cgt}
\bea \la{ad-abv}
&& a_1=a, \quad a_2=a^* \sigma^2, \quad b_1=b, 
\quad b_2=b^* \sigma^2, \xx
&& v_1(x)={v_2(x)}^*=v(x)
\eea
and anti-hermitian generators of $U(N)$ as $T^A,\; A=1,\cdots,N^2$
which are normalized as
\be \la{un-gen}
{\rm tr}(T^A T^B)=-\half \delta_{AB}
\ee
where $T^{N^2}=\frac{1}{i\sqrt{2N}} {\bf 1}_N$.
Then the propagator $G^{AB}(x,y)$ for the adjoint representation 
can be obtained by
multiplying Eq.\eq{ad-green} by $T^A_{us},\;T^B_{vt}$ and summing over
$u,s,v,t=1,\cdots,N$:
\bea \la{adgreen-ex}
G_{AB}(x,y) &=&
[{v(x)}^\dagger]_{u,\lambda} T^A_{us} [v(x)]_{\rho,s} 
G^{(0)}(x,y)[v(y)]_{\lambda,v} T^B_{vt} [{v(y)}^\dagger]_{t,\rho} \xx
&&+\frac{1}{4\pi^2} M_{ij,lm} [{w(x)}^\dagger]_{u,i\alpha} 
T^A_{us}[w(x)]_{j\alpha,s}
[w(y)]_{l\beta,v} T^B_{vt} [{w(y)}^\dagger]_{t,m\beta},
\eea
where $\lambda,\rho=1,\cdots,N+2k$ are ADHM indices 
and we introduced a $2k \times N$ matrix $w(x)=b^\dagger v(x)$.

\section{Discussions}

We have explicitly constructed Green functions for a scalar field in 
an arbitrary representation of gauge group 
propagating in noncommutative instanton background. 
We have shown that the propagators in the adjoint representation 
are deformed by noncommutativity while 
those in the fundamental representation 
have exactly the same form as the commutative case.

We showed, generalizing the argument in \ct{bccl} to noncommutative
space, that the propagators for spinor and vector fields can be 
constructed in terms of those for the scalar field in noncommutative 
instanton background. However it was pointed out in \ct{bccl} that the
vector propagator suffers from an infrared divergence. 
Let's discuss this problem in our context. The vector propagator can
be constructed by the operator expression \eq{op-G} which is involved
to the convolution integral over $z$-coordinates
\be \la{convolution}
\langle x|\Bigl(\frac{1}{D^2}\Bigr)^2_{AB} |y\rangle=\Tr_\CH^z 
G_{AC}(x,z)G_{CB}(z,y),
\ee  
where $G_{AB}(x,y)$ is defined by \eq{adgreen-ex}.
Using the asymptotic behavior \ct{cg,dhkmv} of 
several ADHM quantities at large $z$ limit, 
in which the noncommutativity of
space is irrelevant, one can check that the integral \eq{convolution}
is logarithmically divergent or logarithmically divergent sum,
e.g. $\sum_n n^{-1}$, in noncommutative case. According to \ct{bccl},
one can see that this divergence is coming from the zero mode fluctuations
corresponding to global gauge rotations and an overall scale change,
which have already been contained into the zero mode sum \eq{ym-q}, 
where the zero mode for the scale change is generated 
by the Lorentz rotation of the gauge zero modes. 
However note that the global gauge
zero modes are not normalizable on ${\bf R}^4$ or $\do$ since it is
noncompact space \ct{bcgw,csw}. 
A natural way to remove this logarithmic divergence
is to put the theory on compactified Euclidean space, 
i.e. ${\bf S}^4$, as in \ct{jr-c,nisc}. 
As shown in \ct{nisc}, this compactified Euclidean formalism 
provides a gauge invariant normalization for the global gauge zero modes 
since the volume of ${\bf S}^4$ is now finite. 
Thus the divergence in the vector propagator may be cured in this way 
because the convolution integral \eq{convolution} on ${\bf S}^4$ 
can be finite. It will be interesting to see 
if the infrared divergence in the
vector propagator can be cured by an ``appropriate'' compactification 
of noncommutative space in the same way.

Let's consider the massless Dirac equation 
defined by the covariant derivative \eq{der-2f} in the background of
anti-self-dual instantons. 
In this case it has only positive chirality
solutions \ct{our} described by two spinors $\psi_R = \psi_{us,\alpha}$ 
satisfying
\be \la{dirac-2f}
\sigma^\mu D_\mu \psi_R=0.
\ee
Take the same ansatz as the commutative case \ct{cgt} 
\be \la{zero-ten}
\psi_{us,\alpha}=[v_1(x)^\dagger b_1 \sigma^2 f_1(x)]_{u,i\alpha}
[v_2(x)^\dagger d]_{s,i} +
[v_1(x)^\dagger c]_{u,i}[v_2(x)^\dagger b_2 \sigma^2 f_2(x)]_{s,i\alpha},
\ee
where $c,d$ are constant $(N_1+2k_1) \times k_2$ and 
$(N_2+2k_2) \times k_1$ matrices to be determined. 
Here we are using ordinary multiplication rather than 
the tensor product \eq{multi-law} since $c, d$ are coupling two spaces
$G_1$ and $G_2$ together.
In the case of adjoint
representation, using \eq{ad-abv}, the ansatz \eq{zero-ten} can be 
arranged into the form
\be \la{adzero-ten}
\psi_R = v(x)^\dagger \CM f(x) b^\dagger v(x)
- v(x)^\dagger b f(x) \CM^\dagger v(x),
\ee
with $(N+2k) \times k$ matrix $\CM$.  
Using the formula \eq{adhm-formula} and
\be \la{formu}
{\bar \sigma}^\mu \Bigl(\Delta_a(x)^\dagger b_\a \Bigr) 
{\bar \sigma}^\mu = -2 b_a^\dagger \Delta_a(x), 
\qquad \sigma^\mu \Bigl(b_a^\dagger \Delta_a(x) \Bigr) 
{\bar \sigma}^\mu = 2\Bigl(\Delta_a(x)^\dagger b_a+b_a^\dagger 
\Delta_a(x) \Bigr), 
\ee
it is straightforward to calculate Eq.\eq{dirac-2f}
\bea \la{dirac-cal}
\sigma^\mu D_\mu \psi_R &=& 2\vare_{\beta\gamma}
[v_1(x)^\dagger b_1 f_1(x)]_{u,i\beta} \xx
&& \Bigl([v_2(x)^\dagger b_2 f_2(x)]_{s,j\gamma} 
[\Delta_2(x)^\dagger d]_{j\alpha,i}
-[\Delta_1(x)^\dagger c]_{i\alpha,j}
[v_2(x)^\dagger b_2 f_2(x)]_{s,j\gamma} \Bigr).
\eea

In the commutative case \ct{cgt}, 
the equation \eq{dirac-2f} requires that
\be \la{com-sol}
[a_1^\dagger c]_{i\alpha,j}
=[a_2^\dagger d]_{j\alpha,i}, \qquad [b_1^\dagger c]_{i\alpha,j}
=[b_2^\dagger d]_{j\alpha,i}.
\ee
However, in the noncommutative space, we cannot say that the solution of
\eq{dirac-2f} would be \eq{com-sol} since ${\bf x}$ does not necessarily
commute with $v_2(x)^\dagger b_2 f_2(x)$. 
So the simple minded ansatz \eq{zero-ten} 
doesn't work for the noncommutative space. 
To find the fermionic zero modes for
the gauge group $G_1 \times G_2$, it may be needed to apply 
a systematic way for the tensor
product of instantons as done in \ct{cgt}. Unfortunately, the ADHM
construction for the tensor product involves in tedious and
complicated manipulations even for commutative space. 
We didn't succeed the generalization to the noncommutative space yet. 
Anyway we put this problem for future work.

In Sec. 4, we observed that the $x$-dependent matrix $M$ \eq{c-cal2}
is equal to the constant matrix $M$ \eq{matrix-M} and the
matrix $M$ is deformed by the noncommutativity only for non-BPS
instantons. In commutative space, 
the $x$-independence of the matrix $M$ is 
a result of conformal invariance and the conformal invariance 
has an important role to calculate multi-instanton determinant 
\ct{cgot,osborn,bl,jack}. 
Since, for BPS instantons (commutative instantons are always BPS), 
the matrix $M$ has the same form as the
commutative one, the conformal invariance 
for this background should be manifest. 
Although the matrix $M$ is deformed by the noncommutativity for non-BPS
instantons, it is still a constant matrix. 
Thus one may expect (deformed) conformal invariance 
even for the non-BPS background. In \ct{our}, we observed that the
conformal zero modes have a similar deformation by the noncommutativity 
and we speculated that the conformal symmetry has to act nontrivially 
only on the $SU(N)$ instanton sector. 
These deformations of conformal symmetry in zero modes and
propagators should be related to each other.

Let's briefly discuss the conformal property of 
the matrix $M$ \eq{matrix-M} in the
BPS background, in which the $\sigma_{\mu\nu} \theta^{\mu\nu}$ 
term vanishes. From Eq.\eq{c-cal2} and Eq.\eq{matrix-M}, 
we see that the matrix $M$
in \eq{matrix-M} is invariant under the transformations
\be \la{conf-tr1}
a_a \to a_a + b_a {\bf \bar p},\quad b_a \to b_a, \quad a=1,2.
\ee
Since it is symmetric under interchange of $a_a$ and $b_a$, 
it is also invariant under the
transformations (use $\sigma^\mu\sigma^\nu + 
{\bar \sigma^\mu}{\bar \sigma}^\nu ={\rm tr}_2 (\sigma^\mu\sigma^\nu)=
{\rm tr}_2({\bar \sigma^\mu}{\bar \sigma}^\nu)$ to check)
\be \la{conf-tr2} 
a_a \to a_a, \quad b_a \to b_a + a_a {\bf \bar p}.
\ee 
While, under the transformations
\be \la{conf-tr3}
a_a \to a_a {\bf \bar p}, \quad b_a \to b_a {\bf \bar q},
\ee
it changes by a factor $p^2q^2$ for any quaternions ${\bf \bar p}, 
{\bf \bar q}$ (use \eq{sigma} to check). 
This factor can be scaled to unity in terms of 
simultaneous global scaling of $a_a,
b_a$ by a real number.
The above transformations \eq{conf-tr1}-\eq{conf-tr3} actually
correspond to unimodular conformal group \ct{cgt} 
(so fifteen parameter group).

If we try to generalize the above consideration to the non-BPS
instantons, in which we have 
the $\sigma_{\mu\nu} \theta^{\mu\nu}$ term, 
we immediately meet some nontrivial problems. 
The main source of this difficulty is that 
the matrix $M$ \eq{matrix-M} is asymmetric 
under interchange of $a_a$ and $b_a$ due to the presence of the
inhomogeneous term, i.e. $\sigma_{\mu\nu} \theta^{\mu\nu}$.
The scale transformation \eq{conf-tr3} 
doesn't generate an overall scale either 
because of the inhomogeneous term, 
thus some modified transformation would be genuinely required.
At current stage, we don't know how to modify the conformal
transformations \eq{conf-tr1}-\eq{conf-tr3}. 
We hope to report some progress along this line in near future.


\section*{Acknowledgments}
We thank Keun-Young Kim for helpful discussions and careful reading of
this manuscript. BHL is supported by the Ministry of Education, BK21
Project No. D-0055 and by grant No. R01-1999-00018 from the
Interdisciplinary Research Program of the KOSEF. HSY is supported
by NSC (NSC90-2811-M-002-019). He also acknowledges NCTS as well as
CTP at NTU for partial support.

\newpage


\nc{\np}[3]{Nucl. Phys. {\bf B#1}, #2 (#3)}

\nc{\pl}[3]{Phys. Lett. {\bf B#1}, #2 (#3)}

\nc{\prl}[3]{Phys. Rev. Lett. {\bf #1}, #2 (#3)}

\nc{\prd}[3]{Phys. Rev. {\bf D#1}, #2 (#3)}

\nc{\ap}[3]{Ann. Phys. {\bf #1}, #2 (#3)}

\nc{\prep}[3]{Phys. Rep. {\bf #1}, #2 (#3)}

\nc{\ptp}[3]{Prog. Theor. Phys. {\bf #1}, #2 (#3)}

\nc{\rmp}[3]{Rev. Mod. Phys. {\bf #1}, #2 (#3)}

\nc{\cmp}[3]{Comm. Math. Phys. {\bf #1}, #2 (#3)}

\nc{\mpl}[3]{Mod. Phys. Lett. {\bf #1}, #2 (#3)}

\nc{\cqg}[3]{Class. Quant. Grav. {\bf #1}, #2 (#3)}

\nc{\jhep}[3]{J. High Energy Phys. {\bf #1}, #2 (#3)}

\nc{\hep}[1]{{\tt hep-th/{#1}}}


\end{document}